\documentclass[11pt,fleqn]{article}

\usepackage{amssymb,amsmath}
\usepackage{appendix}
\usepackage{graphicx}
\usepackage{caption}
\usepackage{color}		% Need the color package
\usepackage{epsfig}
\usepackage{float}

\newcommand{\be}{\begin{equation}}

\newcommand{\ee}{\end{equation}}
\newcommand{\reg}{\ensuremath{ \Sigma}}
\newcommand{\s}{\ensuremath{\partial \Sigma}}
\newcommand{\hrb}{\ensuremath{\mathcal{H}^B_\Sigma}}
\newcommand{\hrs}{\ensuremath{\mathcal{H}^S_\Sigma}}
\newcommand{\proj}{{\mathbb{P}}}
\newcommand{\id}{{\mathcal{I}}}

\begin{document}
\begin{titlepage}
\bigskip
%\rightline{hep-th/yymmnnn}

\bigskip\bigskip\bigskip\bigskip

\centerline{\Large \bf {Properties of Causal Holographic Information}}

\bigskip\bigskip
\bigskip\bigskip

\centerline{\bf Ben Freivogel$^{1,2}$ and Benjamin Mosk$^1$}
\medskip
\centerline{\small $^1$ ITFA,
Universiteit van Amsterdam, Amsterdam, the Netherlands}
\centerline{\small $^2$ GRAPPA,
Universiteit van Amsterdam, Amsterdam, the Netherlands}

 \begin{abstract} 
Causal holographic information [1] is a variant of the Ryu-Takayanagi proposal for the entanglement entropy of a spatial region in the context of AdS/CFT, but with the bulk surface defined by causality rather than extremality. We investigate the properties of causal holographic information, focusing in particular on the universal coefficient of the logarithmically divergent term. We find that this coefficient contains a novel conformal invariant that cannot be written as an integral of local quantities. By considering higher curvature corrections in the bulk, we identify the coefficient of the $a$ and $c$ central charges in 4 dimensions. Finally, we speculate about which CFT quantity could correspond to the causal holographic information.
\end{abstract}

\end{titlepage}

\tableofcontents

\newpage
\section{Introduction}
Despite the remarkable success of the AdS/CFT correspondence, basic aspects of the map between bulk and boundary remain elusive. In trying to understand the emergence of locality in the bulk, it is natural to ask which CFT degrees of freedom describe a particular region in the bulk. Recently, there have been intriguing hints of a simple correspondence between $geometric$ subregions of the boundary and geometric subregions of the bulk.

Suppose that we divide a spatial slice of the CFT into two parts, \reg\ and its complement, across a surface \s.  There are two natural bulk regions associated to the CFT subregion \reg. The first is bounded by the bulk surface ending on \s\ with extremal area; \cite{rt, hrt} conjectured that the area of this surface in Planck units computes the entanglement entropy of the boundary subregion \reg.  

The success of this proposal  hints at a very simple mapping: dividing the CFT across \s\ corresponds to dividing the bulk across the extremal bulk surface ending on \s\ \cite{VanRaamsdonk:2009ar, VanRaamsdonk:2010pw, chm}. In other words, a simple conjecture is that the CFT restricted to \reg\ describes the bulk region inside the extremal surface ending on \s.

The second natural bulk region associated to \reg\   is smaller: it is the bulk ``causal wedge" associated to \reg, which we define precisely below.  \cite{blr, Czech:2012bh, hr} suggested that this bulk wedge is the smallest bulk region that must be described by the CFT in the subregion \reg, because it is the region of the bulk that can be directly probed by local observables in the causal development of \reg.\footnote{\cite{bflrz} argued that in fact nonlocal operators are needed even to describe this causal wedge, but we leave such detailed discussions aside in our present work.} 
\begin{figure}[h]
\makebox[\textwidth]{
\scalebox{0.4}{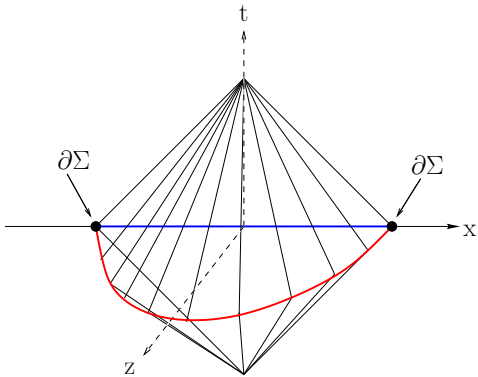  } 
}
\caption{Visualization of the causal information surface $\Xi_{\Sigma}$ (red) reaching into the bulk (z-direction), constructed from the causal diamond of $\Sigma$. The area $\Sigma$ is indicated by the blue line.}
\end{figure}

The natural idea is that the entire bulk region bounded by the $extremal$ surface corresponds to the full CFT  density matrix $\rho_\reg$ of the subregion \reg, while the bulk causal wedge corresponds to a modified density matrix, $\tilde \rho_\reg$, that retains only this information available to local probes.
  
  The bulk region within the causal wedge is bounded by the ``causal information surface."
 Hubeny and Rangamani \cite{hr} focused attention on the area of this surface in Planck units, which they called the causal holographic information, $\chi$. 
   As argued recently by Bianchi and Myers \cite{Bianchi:2012ev}, the entanglement entropy of a spacetime region is given at leading order in $G$ by
$
S = A/4G+ ...
$
where $A$ is the area of the entangling surface. Therefore, from the bulk point of view it seems that causal holographic information is the entanglement entropy of the causal wedge at leading order in $G$.
%  There is some information contained in \reg, for example in large Wilson loops, that is not accessible to any local observer within the causal development of \reg. A natural idea is to start from the full density matrix $\rho_A$ for the region A, and throw away this nonlocal information, giving a new density matrix $\tilde{\rho}_A$. 
In CFT terms,  the causal holographic information is the von Neumann entropy of the modified density matrix $\tilde{\rho}_\reg$,
\be
\chi = S_{\rm ent}(\tilde{\rho}_\reg)
\ee
to leading order in $N$.
We will suggest more precisely, though not definitively, how to construct $\tilde{\rho}_A$ in section 6.

%The second surface was defined recently by Hubeny and Rangamani \cite{hr}. It is the boundary of the bulk ``causal wedge" associated to \reg, as we will define more precisely below. The area of this surface is called the causal holographic information, $\chi$, and it remains unclear what CFT quantity it computes. We first derive some properties of causal holographic information, then speculate about its meaning.

Our main focus, however,  is on computing the properties of causal holographic information. 
In a 4-dimensional boundary theory both the entanglement entropy and the causal holographic information are quadratically UV divergent and regulator dependent.
 However, there is a subleading universal logarithmically divergent term whose coefficient  is independent of both the state and the regulator. 

For the entanglement entropy, this universal constant is a linear combination of the central charges in the CFT and also depends on the geometry of the entangling surface. An explicit expression for this universal coefficient in $d=4$ boundary dimensions has been derived \cite{solodukhin}. For the case of a flat boundary geometry, the entanglement entropy is
\begin{equation} \label{solo}
 \begin{aligned}
%&A\frac{\pi}{8}\int_{\partial \Sigma}\left(R_{abab}-2R_{aa}+R-K_{\mu \nu}^aK^{a \mu\nu}+K_aK^a \right) \\
%-&B\frac{\pi}{8}\int_{\partial \Sigma}\left(R_{abab} -R_{aa}+\frac{R}{3} - K_{\mu \nu}^aK^{a \mu\nu}+\frac{1}{2}K_aK^a\right) \\
S = \dots
 +\log \frac{1}{\epsilon} \left[ -\frac{a}{720\pi}\int_{\partial \Sigma} R_{\partial \Sigma} +
\frac{c}{720 \pi}\int_{\partial \Sigma}\left( - K_{\mu \nu}^aK^{a \mu\nu}+\frac{1}{2}K_aK^a\right) \right] + \dots
 \end{aligned}
\end{equation}
where $R$ and $K$ are the intrinsic and extrinsic curvatures, and $a$ and $c$ are the central charges of the CFT. Note that this expression is an integral over the entangling surface \s\ of local geometric quantities, such as extrinsic and intrinsic curvatures.

We compute the corresponding expansion for causal holographic information for a region with arbitrary shape within a flat boundary geometry. We find a simple formula for the logarithmically divergent term, but now it is $not$ the integral of local, geometric quantities on the boundary. For the case that the bounding surface \s\ lies on a constant time slice, we obtain
\begin{equation}
 \begin{aligned}
\chi = \dots +\log \frac{1}{\epsilon} \left[ -\frac{a}{720\pi}\int d^2\xi \sqrt{\tilde{g}}R_{\partial \Sigma} + \frac{c}{720\pi}\int d^2\xi \sqrt{\tilde{g}}\left(R_{\partial \Sigma}+2\left(\frac{1}{\tau^2}+\frac{K}{\tau} \right) \right)\right] + \dots
 \end{aligned}
\end{equation}
The quantity $\tau$ appearing here is the ``height" of the causal diamond associated to \reg\ at that location, which we will define more precisely later.  But note that the value of $\tau$ at some point on \s\ cannot be determined from the local geometry; for example, if the region \reg\ is a strip, then $\tau$ is determined by the width of the strip. Furthermore, as we argue, this coefficient is invariant under Weyl transformations of the boundary metric.

The paper is organized as follows. In section \ref{definitie} we review the definition of causal holographic information. In section \ref{findlogterm}  and in appendices \ref{tauproof} and \ref{lambdaproof} we derive an expression for causal holographic information in the case of a 4-dimensional flat boundary. In section \ref{univers} we show that the log-term must be universal. In section \ref{gaussmod} we discuss a natural modification to the causal holographic information formula and we derive an expression for the 4-dimensional flat boundary case with a Gauss Bonnet gravity dual. In section \ref{speculate} we discuss properties of causal holographic information in relation with a possible CFT quantity that it might represent.

\section{Definition of causal holographic information}\label{definitie}

Given a co-dimension one spacelike surface $\Sigma$ on the boundary $\partial M$ of an asymptotically AdS-spacetime $M$, one can relate two covariantly defined co-dimension two surfaces in the bulk $M$. The first surface is the extremal surface that ends on the boundary of $\Sigma$, $\partial \Sigma$. The surface area is divergent, but can be regulated. The regulated surface area in Planck units of the extremal surface is proposed to be the dual quantity of the entanglement entropy \cite{hrt}:

\begin{equation} \label{ee}
 \begin{aligned}
S_{\Sigma} &= \frac{A_{extremal}}{4G}
 \end{aligned}.
\end{equation}

The second covariantly defined co-dimension two bulk surface is the causal information surface. First we define the boundary causal diamond of $\Sigma$; it is the union of the boundary future and past domains of dependence:
\begin{equation}
 \begin{aligned}
\Diamond_{\Sigma} &= D^+_{\partial M}(\Sigma) \cup D^-_{\partial M}(\Sigma)
 \end{aligned}.
\end{equation}

\begin{figure}[H]
\makebox[\textwidth]{
\scalebox{0.4}{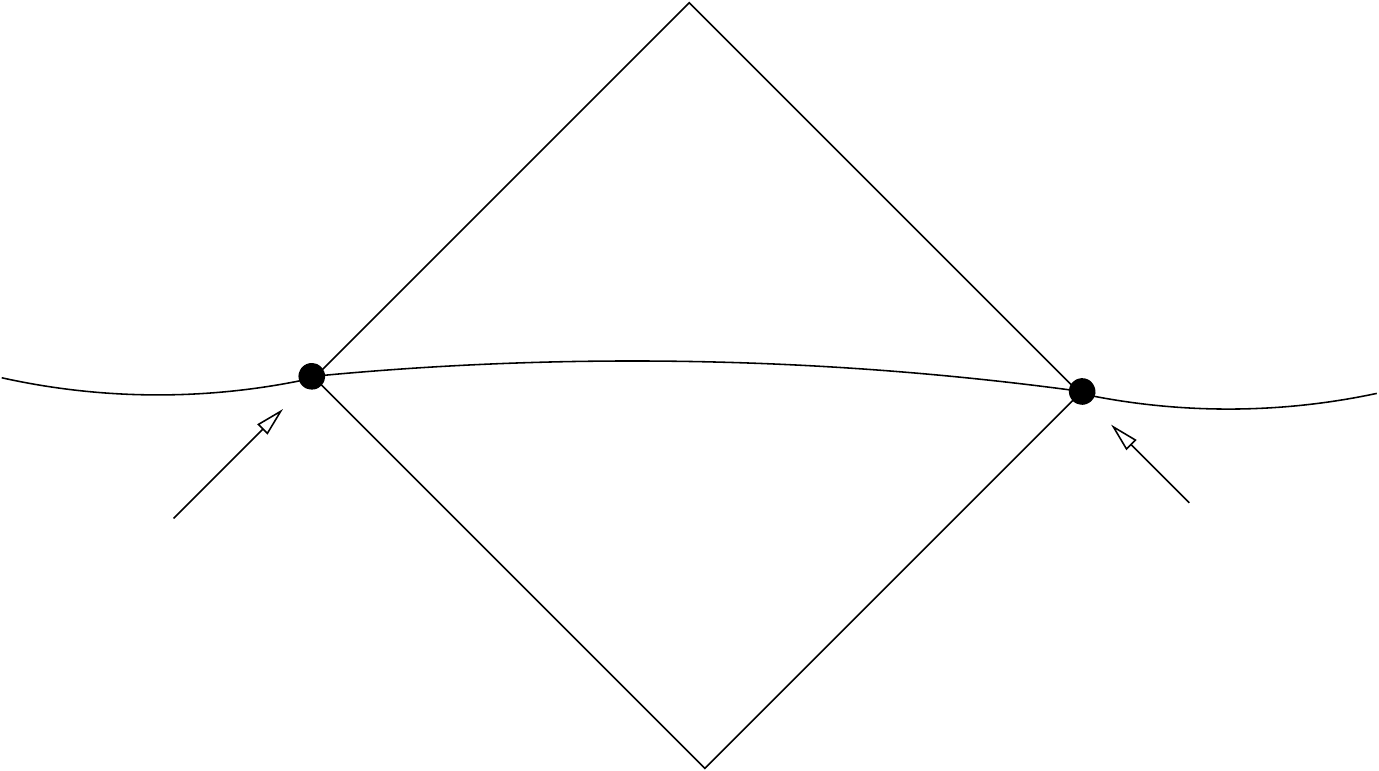  } 
\scalebox{0.4}{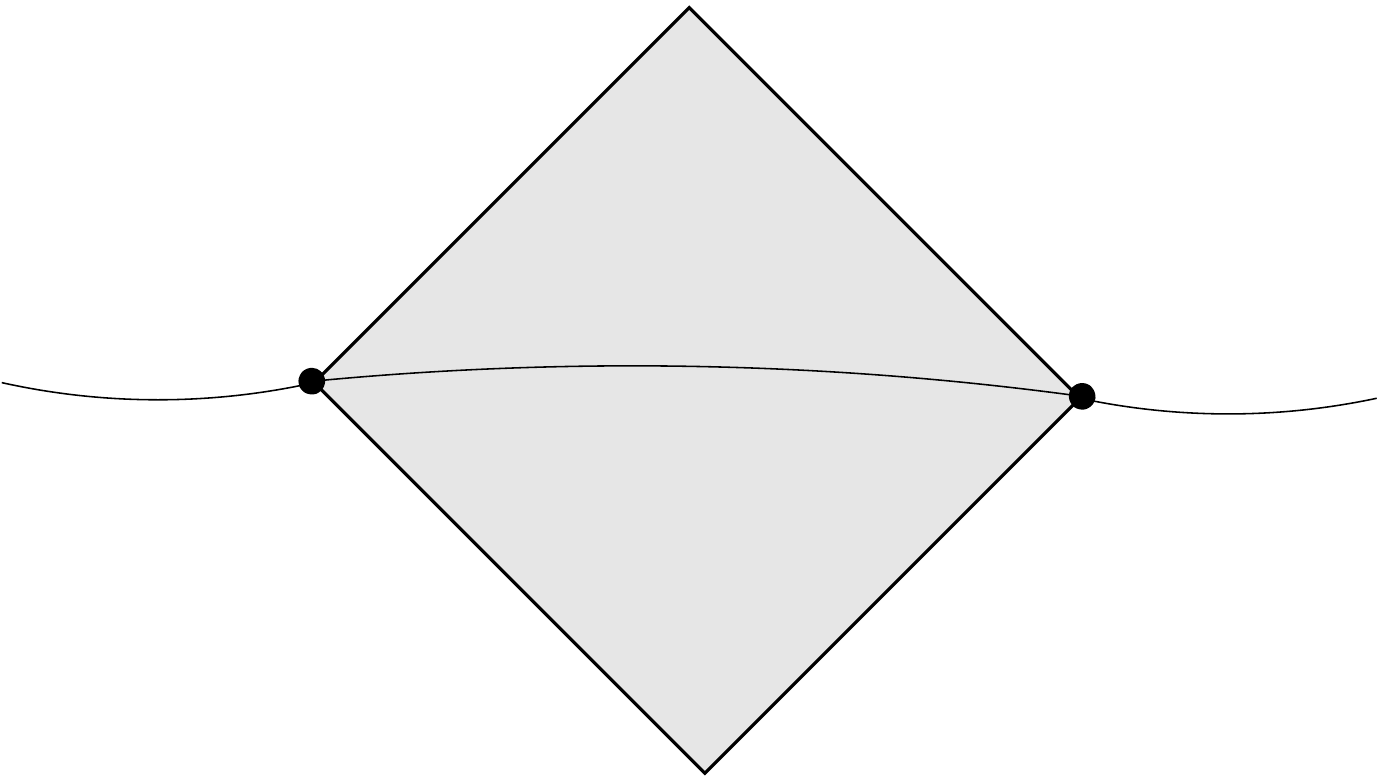  } 
}
\caption{Visualization of the region $\Sigma$ and the boundary $\partial \Sigma$ (left) on a (1+1)-dimensional boundary and the construction of the causal diamond $\Diamond_{\Sigma}$  (shaded area) consisting of the union of the future and past domains of dependence $D^{+}_{\partial M}\left(\Sigma \right)$ and $D^{-}_{\partial M}\left(\Sigma \right)$ (right).}
\end{figure}

The causal diamond $\Diamond_{\Sigma}$ of $\Sigma$ is fully determined by $\partial \Sigma$. We will refer to the points on top and on the bottom of the causal diamond as $\mathcal{C}^{\pm}$ respectively. To be more precise, $\mathcal{C}^+$ is the set of points $p$ in $\Diamond_{\Sigma}$ such that the intersection between the future lightcone of $p$ and $\Diamond_{\Sigma}$ only includes $p$ itself. We will refer to points in $\mathcal{C}^{\pm}$ as future and past ``caustics''.

\begin{figure}[H]
\makebox[\textwidth]{
\scalebox{0.6}{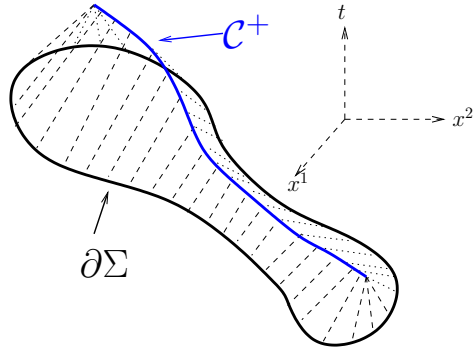  }
}
\caption{Visualization of the boundary causal diamond $\Diamond_{\Sigma}$ in case of a (2+1)-dimensional boundary. The set of future ``caustics'' $\mathcal{C}^+$ is indicated by the blue line.}
\end{figure}

The so called causal wedge $\blacklozenge_{\Sigma}$ of $\Sigma$ is the intersection of the bulk future and past domains of influence of $\Sigma$:

\begin{equation} \label{wedge}
 \begin{aligned}
\blacklozenge_{\Sigma} &= J_M^+(\Diamond_{\Sigma}) \cap J_M^-(\Diamond_{\Sigma}).
 \end{aligned}
\end{equation} 

The co-dimension two bulk surface we are interested in is the intersection of the bulk boundaries of the bulk future and past domains of influence of the causal diamond $\Diamond_{\Sigma}$: 
\begin{equation} \label{cis}
 \begin{aligned}
\Xi_{\Sigma} &= \partial J_M^+(\Diamond_{\Sigma}) \cap \partial J_M^-(\Diamond_{\Sigma}).
 \end{aligned}
\end{equation}

More elucidating graphics of the causal information surface $\Xi_{\Sigma}$ can be found in \cite{hr}.
Causal holographic information is defined in a way similar to the entanglement entropy (\ref{ee}):
\begin{equation}
 \begin{aligned}
\chi_{\Sigma} &= \frac{A_{\Xi_{\Sigma}}}{4G},
 \end{aligned}
\end{equation}
where $\Xi_{\Sigma}$ is the causal information surface as defined in (\ref{cis}) \cite{hr}.

The causal information surface has a surface area that is in general larger than or equal to the surface area of the extremal surface, when both surfaces are regularized in a consistent way. If the bulk spacetime is static and one picks $\Sigma$ to be on a constant time slice, the extremal surface and the causal information surface both lie in the same bulk constant time slice. In this case the causal information surface either coincides with the minimal surface or it is located closer to the boundary \cite{hr}.

Causal holographic information does not satisfy the strong subadditivity property \cite{hr} whereas the holographic entanglement entropy does \cite{Headrick:2007km}. Also, for a pure state we can find examples with $\chi_{\Sigma} \neq \chi_{\Sigma^c}$ \cite{hr}, where $\Sigma^c$ is the boundary complement of $\Sigma$.

\section{Finding an expression for the universal log-term}\label{findlogterm}
\subsection{The strip: example of a nonlocal log-term}
The example of the strip in a 4-dimensional flat background shows that the coefficient of the logarithmically divergent term of causal holographic information cannot be expressed as just an integral over $\partial \Sigma$ of local geometric quantities. Using coordinates $(x^1,x^2,x^3,t)$ for the flat boundary, the strip is the area $\Sigma_{strip}$ defined by:

\begin{equation}
 \begin{aligned}
 -\frac{w}{2} \leq x^1 \leq \frac{w}{2}, \ \ \ \ \ t=0 ,
 \end{aligned}
\end{equation}
with $x^2$ and  $x^3$ unconstrained.

The coefficient of the logarithmically divergent term of the causal holographic information for the strip (d=4) is proportional to $\frac{L^2}{w^2}$ \cite{hr}, where $L$ is an IR-regulator in the directions $x^2$ and $x^3$. For the strip there is no local geometric quantity on $\partial \Sigma$ (the two plates at $x^1 = \pm \frac{w}{2}$) that depends on the separation distance $w$. This means that we cannot write the coefficient of the logarithmically divergent term in $\chi_{\Sigma}$ as the integral of a local quantity, which is possible for the coefficient of the logarithmically divergent term in entanglement entropy (\ref{solo}).

\subsection{Caustics of the boundary causal diamond}
The causal diamond $ \Diamond_{\Sigma}$ is generated by null rays emanating normally from $\partial \Sigma$ \cite{hr}. For each point $x(\xi)$ on $\partial \Sigma$ there are two unique null normal vectors (up to a sign). The inward-pointing null geodesics that emanate orthogonally from this point terminate in past and future caustics $x_{\vee} \in \mathcal{C}^-$ and $x_{\wedge} \in \mathcal{C}^+$ respectively. Given a point on $\partial \Sigma$, we need information about other points on $\partial \Sigma$ to determine where the null geodesics will intersect for the first time with null geodesics emanating from another point on $\partial \Sigma$. The location of this intersection point is determined by one or more other points on $\partial \Sigma$.

\subsection{Entering the bulk}
We focus in our discussion on the case of $d=4$ boundary dimensions; however, a logarithmically divergent term can be present in the causal holographic information whenever the number of boundary dimensions is even. For $d$ even and $d>2$ the asymptotically $AdS$-metric can be put in Fefferman Graham form \cite{fgexp} \cite{feffermangraham2}:
\begin{equation}
 \begin{aligned}
ds^2 &= G_{MN}dX^MdX^N \\
&= \frac{dz^2}{z^2}+\frac{1}{z^2}g_{\mu \nu}(z,x)dx^{\mu}dx^{\nu} \\
g_{\mu \nu}(z,x) &=  g^{(0)}_{\mu \nu}(x)+z^2  g ^{(2)}_{\mu \nu}(x) +...+ z^d \left(g^{(d)}_{\mu \nu}(x)+\log{z} \ h_{\mu \nu}(x) \right)+O(z^{d+1}).
 \end{aligned}
\end{equation}
Here $g^{(0)}_{\mu \nu}(x)$ is the boundary metric, and $g^{(2)}_{\mu \nu}$ and $h_{\mu \nu}$ are determined by $g^{(0)}_{\mu \nu}$, but $g^{(4)}_{\mu \nu}$ can be chosen independently; it encodes information about the state of the theory.

Close to the boundary, points on the causal information surface $\Xi_{\Sigma}$ can be mapped to points on $\partial \Sigma$. 
We expand the embedding function close to the boundary around points on $\partial \Sigma$:
\begin{equation} \label{inbulk}
 \begin{aligned}
x^{\mu}_{bulk}(z,\xi_1,\xi_2) &= x^{\mu}_{boundary}(\xi_1,\xi_2)+y^{\mu}(\xi_1,\xi_2)z^2+O(z^4).
 \end{aligned}
\end{equation}

Generally, the component of $y^{\mu}$ that is tangent to $\partial \Sigma$ does not give a contribution to the logarithmically diverging term in the area of the bulk surface. We can express the normal component of $y^{\mu}$ in terms of the boundary normal vectors $\{ N_a^{\mu}\}$ of $\partial \Sigma$: 
\begin{equation} \label{ynormal}
 \begin{aligned}
y^{\mu} &= y^{\mu}_{\parallel} + y^{\mu}_{\perp} &&= y^{\mu}_{\parallel} + \lambda^{a}N_{a}^{\mu}.
 \end{aligned}
\end{equation}

For any such surface, the surface area that is regulated by a radial $z=\epsilon$ cutoff for an asymptotically AdS-space is:
\begin{equation}\label{normalarea}
 \begin{aligned}
A &= \frac{L_{AdS}^3}{2} \frac{A_{\partial \Sigma}}{\epsilon^2} \\
&+ \log \left(\frac{1}{\epsilon} \right) \frac{L_{AdS}^3}{2}
\int d^2\xi \sqrt{\tilde{g}}
\left( 4\lambda^a \lambda^b p_{ab}+ 2\lambda^a K_a 
 + h^{\mu \nu} g^{(2)}_{\mu \nu}
\right) \\
&+\text{finite},
 \end{aligned}
\end{equation}
where $N_{a}^{\mu}N_{b}^{\nu}g^{(0)}_{\mu\nu} = p_{ab}$, $h_{\mu \nu} = g^{(0)}_{\mu \nu}-N^a_{\mu}N^b_{\nu}p^{ab}$, $K^a$ is the trace of the extrinsic curvature, and an expression for $g^{(2)}_{\mu \nu}$ is given later (\ref{g2}). 

\subsection{Flat boundary case}
For entanglement entropy, the normal component of $y$ (\ref{inbulk},\ref{ynormal}) is fixed by the extremality condition to $\lambda^a = -\frac{K^a}{4}$. For the causal information surface the $\lambda_a(x)$'s in the expansion (\ref{inbulk},\ref{ynormal}) do not just depend on local geometric quantities of $\partial \Sigma$, as can be seen in the example of the strip.

We first analyze the simpler case where the entangling surface lies on a constant time slice, then the more general case of an arbitrary spacelike surface embedded in flat Minkowski spacetime.

\subsubsection{Constant time slice case} 
In the case of a static spacetime with a flat boundary, the analysis is easier if we pick the surface $\partial \Sigma$ to be on a constant time slice. Boundary normal null geodesics emanating from $\partial \Sigma$ are simply straight null rays proportional to the null normal vectors of $\partial \Sigma$. When $\partial \Sigma$ is on a constant time slice $t=0$, the causal information surface should also be on the $t=0$ slice by symmetry considerations; both the bulk causal wedge and the boundary causal diamond are symmetric in the $t=0$ plane. Points on the causal information surface are on the bulk boundaries of the bulk future and past domains of influence of the boundary causal diamond $\Diamond_{\Sigma}$ (\ref{cis}). 

The object $\Sigma$ on the boundary will ``shrink'' as one moves deeper into the bulk along the radial coordinate. In appendix (\ref{tauproof}) we determine how this happens as a function of the radial Poincar\'e coordinate $z$. 
%Taking $\partial \Sigma$ to be on a constant time slice simplifies the analysis since for any point $p$ in $\Diamond_{\Sigma}$ on the $t=0$ slice the past and future caustics on  $\Diamond_{\Sigma}$ that extremize the distance to $p$ are related to the same point(s) on $\partial \Sigma$. In the more general case where $\Sigma$ is not on a constant timeslice the analysis is complicated since then the past and future caustics on  $\Diamond_{\Sigma}$ that extremize the distance to $p$ might not be related to the same points on $\partial \Sigma$.

If the future caustic that is separated from $x(\xi)$ on $\partial \Sigma$ by the null normal emanating from $x(\xi)$ is located at $t = \tau$ and the spacelike normal vector in the $t=0$ plane in $x(\xi)$ is given by  $\hat{n}(\xi)$\footnote{We used $\hat{n}$ to be the inward pointing normal together with the conventions $K_{\mu \nu} = + h^{\rho}_{\mu}\nabla_{\rho}N_{\nu}$ and $K = +h^{\mu \nu}K_{\mu \nu}$ for extrinsic curvatures and expansions respectively, where $h$ is an induced metric and $N$ is a normal vector.}, then the embedding function can be written in terms of $\tau$ (Appendix \ref{tauproof}):

\begin{equation}\label{tauexpans}
 \begin{aligned}
 \vec{x}_{bulk}(\xi,z) &= \vec{x}_{\partial \Sigma}(\xi) + \frac{z^2}{2\tau (\xi)}\hat{n}(\xi) - \frac{z^2}{2 \tau (\xi)}b^{\alpha}\vec{T}_{\alpha}(\xi)+O(z^4).
 \end{aligned}
\end{equation}

Expression (\ref{normalarea}) simplifies in this case to
\begin{equation}\label{flat}
 \begin{aligned}
A &= \frac{L_{AdS}^3}{2}\frac{A_{\partial \Sigma}}{\epsilon^2}+\frac{L_{AdS}^3}{2}\int_{\Sigma} d^2\xi \sqrt{g_{\Sigma}}\left( \frac{1}{\tau^2}+\frac{K}{\tau}\right)\log{\frac{1}{\epsilon}} + \text{finite}.
 \end{aligned}
\end{equation}

\begin{figure}[h]
\makebox[\textwidth]{
\scalebox{0.5}{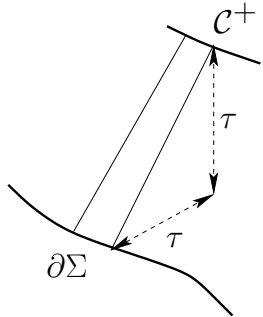  } 
}
\caption{Visualization of the quantity $\tau$. Note that that for each point on $\partial \Sigma$ there is a unique point on $\mathcal{C}^+$ separated by a vector that is proportional to the future directed inward pointing null normal vector. For a surface $\partial \Sigma$ that does not lie on a constant time slice, a single function $\tau$ is not sufficient.
%the boundary causal diamond is no longer symmetric in the $t=0$ plane.
}
\end{figure}

\subsubsection{General spacelike region}
Now dropping the assumption that $\partial \Sigma$ lies on a constant timeslice, but remaining in Minkowski space, the separation in time of a point $p$ on $\partial \Sigma$ and its caustics on the causal diamond may be different for the past and future caustics. 

The past caustic $x^{\mu}_{\wedge}$ and the future caustic $x^{\mu}_{\vee}$ for a point $x^{\mu}(\xi)$ on $\partial \Sigma$ are separated from $x^{\mu}(\xi)$ by null normal vectors:
\begin{equation}\label{lambs}
 \begin{aligned}
x^{\mu}_{\wedge}(\xi) &= x^{\mu}(\xi) + \lambda_{\uparrow}(\xi)N^{\mu}_{\uparrow}(\xi) \\
x^{\mu}_{\vee}(\xi) &= x^{\mu}(\xi) + \lambda_{\downarrow}(\xi)N^{\mu}_{\downarrow}(\xi)
 \end{aligned}
\end{equation}
where $N^{\mu}_{\uparrow}(\xi)$ and $N^{\mu}_{\downarrow}(\xi)$ are the null normal vectors in $x^{\mu}(\xi)$ where we choose to normalize them such that $N^{\mu}_{\uparrow}(\xi)N^{\nu}_{\downarrow}(\xi)\eta_{\mu \nu} = 1$. In appendix (\ref{lambdaproof}) we argue that we can still expand the embedding function similarly to (\ref{tauexpans}):

\begin{equation}\label{lambdaexpans}
 \begin{aligned}
 x^{\mu}_{bulk}(z,\xi) &= x^{\mu}(\xi) + \frac{z^2}{2}\left(\frac{N^{\mu}_{\uparrow}(\xi)}{\lambda_{\downarrow}(\xi)}+\frac{N^{\mu}_{\downarrow}(\xi)}{\lambda_{\uparrow}(\xi)} \right)+\frac{z^2}{2\lambda_{\uparrow}(\xi)}b^{\alpha}_{t}(\xi)T^{\mu}_{\alpha} + O(z^4) \\
 \end{aligned}
\end{equation}

The divergent part of the area is now simply:

\begin{equation}\label{blaat}
 \begin{aligned}
A &= \frac{L_{AdS}^3}{2}\frac{A_{\partial \Sigma}}{\epsilon^2}+ \frac{L_{AdS}^3}{2}\log{\frac{1}{\epsilon}}\int d^2\xi \sqrt{\det g_{\Sigma}}\left(\frac{2}{\lambda_{\uparrow}\lambda_{\downarrow}} + \frac{K_{\uparrow}}{\lambda_{\downarrow}} + \frac{K_{\downarrow}}{\lambda_{\uparrow}} \right)+finite.
 \end{aligned}
\end{equation}
Equation (\ref{lambs}) defines $\lambda_{\uparrow}$ and $\lambda_{\downarrow}$. The null normal vectors can be scaled simultaneously with $\lambda_{\uparrow}$ and $\lambda_{\downarrow}$, but all terms in (\ref{blaat}) are invariant under this rescaling. The absolute value of the distance between the future and past caustics related to a point $x(\xi)$ on $\partial \Sigma$ by (\ref{lambs}) is equal to $\sqrt{2 \lambda_{\uparrow}(\xi) \lambda_{\downarrow}(\xi)}$.

\section{Universality of the log-term}\label{univers}
In this section, we show that the coefficient of the logarithmically divergent term, which we have focused on, is universal. It is independent of the regulator, and also independent of the state. Our result is more general than the particular causal information surface: the same will be true for $any$ covariantly defined surface\footnote{It is important that the shape of the surface near the boundary allows an expansion like  (\ref{inbulk}).} The discussion will take place in a setting with a $d=4$ dimensional boundary, but we believe similar arguments hold for $d = 2n$ with $n>2$. 

\subsection{Regulator independence}
This subsection is to a large extent based on work from Schwimmer and Theisen \cite{Schwimmer:2008yh} \cite{Imbimbo:1999bj}. We briefly review the relevant parts of their work and argue that any covariantly defined co-dimension two bulk surface that can be expanded as in (\ref{inbulk}) yields a universal log-divergence coefficient.

Our strategy for showing the regulator independence is to relate a change in regulator to a change of coordinates in the bulk that leaves the metric in Fefferman Graham form. The regulated area of the surface is computed up to a cutoff value of the new radial coordinate. Such a bulk change of coordinates acts as a conformal transformation on the boundary, so showing the regulator independence of the log term is equivalent to showing that it is Weyl invariant. 

In more detail, to define the regulated area, one first puts the metric in Fefferman Graham form \cite{fgexp} \cite{feffermangraham2}:
\begin{equation}\label{FGform}
ds^2 = G_{MN}dX^MdX^N
= \frac{d\rho^2}{4\rho^2}+\frac{1}{\rho}g_{\mu \nu}(\rho,x)dx^{\mu}dx^{\nu} 
\end{equation}

For $d$ even, we can expand $g_{\mu \nu}(\rho,x)$ as
\begin{equation}\label{FGformg}
 \begin{aligned}
g_{\mu \nu}(\rho,x) &= g^{(0)}_{\mu \nu}(x)+\rho g^{(2)}_{\mu \nu}(x)+...+ \rho^{\frac{d}{2}}\left(g^{(\frac{d}{2})}_{\mu \nu}(x)+\log \rho  \ \ h_{\mu \nu}(x) \right)+O(\rho^{\frac{d}{2}+1}).
 \end{aligned}
\end{equation}

One can take a finite cut off in the radial coordinate $\rho$ in order to regulate the surface area of a bulk surface that ends on $\partial \Sigma$. The ambiguity in choice of cutoff  arises because there is not a unique way to choose coordinates such that the bulk metric is in Fefferman Graham form. 

To see how the area changes, consider an infinitesimal diffeomorphism that leaves the metric in the Fefferman Graham form. Such a transformation is parametrized by a vector field $(\xi^{\rho},\xi^{\mu})$ such that  \cite{Schwimmer:2008yh} \cite{Imbimbo:1999bj}:
\begin{equation}\label{pbhcond}
 \begin{aligned}
\left(\mathcal{L}_{\xi}G\right)_{\rho \mu} &= 0 \\
\left(\mathcal{L}_{\xi}G\right)_{\rho \rho} &= 0.
 \end{aligned}
\end{equation}
Such a PBH-transformation (Penrose-Brown-Hennaux-transformation) corresponds to a Weyl transformation of the boundary metric $g^{(0)}_{\mu \nu}$ and can be parametrized by a function $\omega(x)$. \\

The infinitesimal coordinate transformation 
\begin{equation}\label{coordPHB}
 \begin{aligned}
\rho &= \tilde{\rho}e^{-2\omega(\tilde{x})} \simeq \tilde{\rho} \left(1-2\omega(\tilde{x}) \right) \\
x^{\mu} &= \tilde{x}^{\mu}+a^{\mu}(\tilde{x},\tilde{\rho})
 \end{aligned}
\end{equation}
corresponding to $\xi^{\rho} = -2\omega$ and $\xi^{\mu} = a^{\mu}$ satisfies the requirement (\ref{pbhcond}) and the condition $a^{\mu}(\rho = 0)=0$ if \cite{Imbimbo:1999bj}:
\begin{equation}
 \begin{aligned}
\partial_{\rho}a^{\mu} &= \frac{1}{2}g^{(0)\mu \nu}\partial_{\nu}\omega \\
a^{\mu}(\rho,x) &= \frac{1}{2}\int_0^{\rho} \ d\tilde{\rho} \ g^{\mu \nu}(\tilde{\rho},x)\partial_{\nu}\omega(x) + O(\omega^2)\\
&= \frac{\rho}{2}g^{(0)\mu \nu}(x)\partial_{\nu}\omega(x)+O(\rho^2)+O(\omega^2).
 \end{aligned}
\end{equation}

If the embedding function of the causal information surface can be expanded as in (\ref{inbulk});
\begin{equation}\label{rhoexp}
 \begin{aligned}
x^{\mu}_{bulk}(\rho,\xi_1,\xi_2) &= x^{\mu}_{boundary}(\xi_1,\xi_2)+y^{\mu}(\xi_1,\xi_2)\rho+...,
 \end{aligned}
\end{equation}
we can make use of the covariant definition of the causal information surface by applying the coordinate transformation (\ref{coordPHB}):
\begin{equation}
 \begin{aligned}
\tilde{x}^{\mu}_{bulk}+\frac{\tilde{\rho}}{2}g^{(0)\mu \nu}\partial_{\nu}\omega+...  &= x^{\mu}_{boundary}+y^{\mu}\tilde{\rho}\left(1-2\omega \right)+...
 \end{aligned}
\end{equation}
Collecting powers of $\tilde{\rho}$ we conclude that $y$ transforms in the following way:
\begin{equation}
 \begin{aligned}
\tilde{y}^{\mu} &= y^{\mu}-2\omega y^{\mu}-\frac{1}{2}g^{(0)\mu \nu}(x)\partial_{\nu}\omega(x)+... \\
&\Rightarrow \tilde{y}^{\mu} = e^{-2\omega}\left(y^{\mu}-\frac{1}{2}g^{(0)\mu \nu}\partial_{\nu}\omega \right).
 \end{aligned}
\end{equation}

The normal components $\lambda_a = N_a^{\mu}y^{\nu}g^{(0)}_{\mu \nu}$ of $y$ transform as:
\begin{equation}\label{trans}
 \begin{aligned}
\tilde{\lambda}_a &= e^{-\omega}\left(\lambda_a-\frac{1}{2}N_a\cdot \partial \omega \right).
 \end{aligned}
\end{equation}
The coefficients in the logarithmically diverging terms (\ref{normalarea}) and (\ref{GBresult}) are invariant under such a transformation, given that the other boundary quantities are subject to a conformal transformation. Taking a different cut off does not change these coefficients.

This shows that for $any$ covariantly defined bulk surface with $d=4$ boundary dimensions, the coefficient of the logarithmically divergent term is universal, as we expect from field theory intuition. It would be interesting to ask if there are any additional covariantly defined bulk surfaces associated to a region, since they would define new conformal invariants.

In $d=4$ boundary dimensions (\ref{trans}) can be solved by $\lambda_a = -\frac{K^a}{4}$, where $K^a$ is the trace of the extrinsic curvature for the normal vector $N_a$. This corresponds to the result that is obtained by extremizing (\ref{normalarea}) with respect to $y$. The remarkable thing about the causal information surface is that the logarithmically divergent term cannot be constructed from just local geometric quantities as can be seen in the case of the strip.

Finally, it should be pointed out that one can use freedom to re-parametrize $\partial \Sigma$ to gauge out the tangential part of $y$ \cite{Schwimmer:2008yh}. In fact, we already partially fixed the freedom to re-parametrize the causal information surface $\Xi_{\Sigma}$ by choosing one of the parameters to be $z$ (or equivalently $\rho$).

\subsection{State dependence}
In expansion (\ref{FGformg}), the $g^{(n)}_{\mu \nu}(x)$ with $0<n<d$ are fully determined in terms of $g^{(0)}_{\mu \nu}(x)$ \cite{fgexp} \cite{feffermangraham2}. Terms of order $O(\rho^{\frac{d}{2}})$ do not contribute to the logarithmically diverging term. The coefficient of the logarithmically diverging term is thus state independent since the $g^{(n)}_{\mu \nu}(x)$ with $0<n<d$ do not depend on the state. In the case of a $4$-dimensional boundary only $g^{(0)}_{\mu \nu}(x)$ and 
$g^{(2)}_{\mu \nu}(x)$ are relevant for the coefficient of the logarithmically divergent term. Regardless of the equation of motion for the bulk metric, $g^{(2)}_{\mu \nu}(x)$ is determined by the conformal structure of asymptotically AdS-spacetime:
\begin{equation}\label{g2}
 \begin{aligned}
 g^{(2)}_{\mu \nu}(x) &= -\frac{1}{d-2}\left(R^{(0)}_{\mu \nu} - \frac{R^{(0)}g^{(0)}_{\mu \nu}}{2(d-1)} \right). 
 \end{aligned}
\end{equation}

 \section{Causal Holographic Information with GB-gravity}\label{gaussmod}
For entanglement entropy, the coefficient of the logarithmically diverging term can be written as a linear combination of the central charges $a$ and $c$ (\ref{solo}). One may hope to find a similar expression for the logarithmically diverging term in causal holographic information. In order to investigate this, we extend our analysis to causal surfaces with Gauss-Bonnet gravity in the bulk, where the central charges $a$ and $c$ are distinguishable. 

\subsection{Gauss Bonnet gravity}
The action for Gauss-Bonnet gravity is given by
\begin{equation}
 \begin{aligned}
S &= \frac{1}{16\pi G_N^{(5)}}\int d^5x \sqrt{-g}\left(R+\frac{12}{L^2}+\frac{\lambda L^2}{2}\mathcal{L}_{(2)} \right) \\
\mathcal{L}_{(2)} &= R_{MNRS}R^{MNRS}-4R_{MN}R^{MN}+R^2,
 \end{aligned}
\end{equation}
with negative cosmological constant $\Lambda = -\frac{12}{L^2}$.

There is a stable AdS-solution of the equations of motion with AdS-radius \cite{jan}:
\begin{equation}
 \begin{aligned}
L_{AdS} &= \frac{L}{\sqrt{\alpha}} \\
\alpha &= \frac{2}{1+\sqrt{1-4\lambda}}.
 \end{aligned}
\end{equation}

The central charges in the boundary theory are given by \cite{jan} \cite{Henningson:1998gx} \cite{henningson} \cite{Nojiri:1999mh}:
\begin{equation}\label{central}
 \begin{aligned}
c &= 45\pi \frac{L_{AdS}^3}{G_N^5}\sqrt{1-4\lambda} \\
a &= 45\pi \frac{L_{AdS}^3}{G_N^5}\left(-2+3\sqrt{1-4\lambda} \right).
 \end{aligned}
\end{equation}

\subsection{Modification of the causal information formula}

As argued by \cite{Jacobson:1993vj} \cite{Iyer:1994ys} \cite{clunan}, the formula for the holographic entanglement entropy (\ref{ee}) must be modified for higher derivative gravity. A candidate for the holographic entanglement entropy is obtained by applying the Wald entropy formula to surfaces that end on the entanglement surface $\partial \Sigma$. The entanglement entropy is then given by the surface that extremizes this quantity. 
\begin{equation}\label{Wald}
 \begin{aligned}
S_{Wald} &= -2\pi \int_{horizon} d^{d-1}x \sqrt{h} \frac{\delta \mathcal{L}}{\delta {R^{\mu \nu}}_{\rho \sigma}}\hat{\epsilon}^{\mu \nu}\hat{\epsilon}_{\rho \sigma}
 \end{aligned}
\end{equation}
Applying the Wald entropy formula (\ref{Wald}) to a Killing horizon introduces an ambiguity \cite{Hung:2011xb}. When the Wald entropy formula is applied to a surface that has nonvanishing extrinsic curvature and nonvanishing expansions, this ambiguity is lifted and gives two different candidates that assign an entropy-like quantity to a surface. Hung, Myers, and Smolkin argue that only one of the formulas gives consistent results for holographic entanglement entropy \cite{Hung:2011xb}. We use the formula proposed by Myers, dropping a boundary term since it does not contribute to the log-term:
\begin{equation}\label{GB}
 \begin{aligned}
\chi_{\Sigma} = \frac{1}{4G_N^{(5)}}\int_{\Xi_{\Sigma}} \sqrt{\sigma_{\Xi}}\left(1+\lambda L^2 R_{\Xi} \right) + {\rm boundary\ term},
 \end{aligned}
\end{equation}
where $\sigma_{\Xi}$ is the induced metric on the bulk surface $\Xi$ and $R_{\Xi}$ is the intrinsic curvature.

The intrinsic curvature for a surface in an asymptotically AdS spacetime, for which the embedding function can be expanded as in (\ref{inbulk}) is given by:
\begin{equation}
 \begin{aligned}
R_{\Xi} &= -6+z^2\left(R_{\partial \Sigma}+2h^{\mu \nu}g^{(2)}_{\mu \nu}+4\lambda^aK_a+ 8 \lambda^a\lambda_a\right) + O(z^4),
 \end{aligned}
\end{equation}
where $R_{\partial \Sigma}$, $K^a$ and $h$ are boundary quantities.

The coefficient of the logarithmically diverging term is now proportional to:
\begin{equation}\label{GBresult}
 \begin{aligned}
a \int_{\partial \Sigma}d^2\xi \sqrt{\tilde{g}} \left(-\frac{R_{\partial \Sigma}}{4} \right)+c \int_{\partial \Sigma}d^2\xi \sqrt{\tilde{g}} \left(\frac{R_{\partial \Sigma}}{4} + \frac{1}{2}h^{\mu \nu}g^{(2)}_{\mu \nu}+\lambda^aK_a+2\lambda^a \lambda_a \right).
 \end{aligned}
\end{equation}

We can evaluate this expression for the case of a static spacetime with a flat boundary. When $\partial \Sigma$ is on a constant time slice, we can use (\ref{tauexpans}) in order to get the coefficient of the logarithmically diverging term:

\begin{equation}
 \begin{aligned}
-\frac{a}{720\pi}\int d^2\xi \sqrt{\tilde{g}}R_{\partial \Sigma} + \frac{c}{720\pi}\int d^2\xi \sqrt{\tilde{g}}\left(R_{\partial \Sigma}+2\left(\frac{1}{\tau^2}+\frac{K}{\tau} \right) \right).
 \end{aligned}
\end{equation}

Now considering a spacelike surface $\partial \Sigma$ that is not necessarily on a constant time slice, we can apply (\ref{lambdaexpans}) and find for the coefficient of the logarithmically diverging term:

\begin{equation}
 \begin{aligned}
-\frac{a}{720\pi}\int d^2\xi \sqrt{\tilde{g}}R_{\partial \Sigma} + \frac{c}{720\pi}\int d^2\xi \sqrt{\tilde{g}}\left(R_{\partial \Sigma}+2\left(\frac{2}{\lambda_{\uparrow} \lambda_{\downarrow}}+\frac{K_{\uparrow}}{\lambda_{\downarrow}}+\frac{K_{\downarrow}}{\lambda_{\uparrow}} \right) \right).
 \end{aligned}
\end{equation}

\section{Speculations on the CFT dual of $\chi$}
\label{speculate}
The most important question is what CFT quantity the causal holographic information computes. In this section we will discuss general properties of causal holographic information and suggest different types of field theory quantities that satisfy at least some of these properties. These proposals are speculative and would be subject to further research.

First, causal holographic information for an `entanglement surface' $\partial \Sigma$ behaves in many ways like entanglement entropy:
\begin{itemize}
\item It does not depend on a choice of time slicing. The entire construction proceeds from the boundary \s\ .
\item The leading UV divergence of the causal holographic information for $d>2$ is the same as for the entanglement entropy: the 'area law'-term is proportional to the area of $\partial \Sigma$.
\end{itemize}

On the other hand, Hubeny and Rangamani point out several reasons it {\it cannot} be the entanglement entropy.
\begin{itemize}
\item It does not obey strong subadditivity \cite{hr}.
\item When the entire system is in a pure state, one can violate $\chi_{\Sigma} = \chi_{\Sigma^c}$. 
\end{itemize}

There are also several clues identified by Hubeny and Rangamani:
\begin{itemize}
\item $\chi_{\Sigma} \geq S_{\Sigma}$ (although both are infinite). 
\item $\chi_{\Sigma}$ is `teleological': a source in the future domain of dependence of the region \reg\ can change the causal holographic information of \reg\ .
\item `Mutual causal information' vanishes for separated regions $A$ and $B$: $\chi_A+\chi_B - \chi_{A \cup B} = 0$.
\end{itemize}

 In simple cases, such as a scalar field theory on a lattice, there is a unique way to associate a spatial region to a tensor factor of the whole Hilbert space. The Hilbert space is just a product of the degrees of freedom at each lattice point, so given a region \reg , the full Hilbert space is the tensor product of the degrees of freedom in \reg\ and those in the complement,
\[
\mathcal{H} = \mathcal{H}_\reg \otimes \mathcal{H}_{\reg^c}
\]
where $\reg^c$ is the complement of \reg.
We introduce a lattice to avoid subtleties related to UV divergences, but we do not believe these UV subtleties are relevant to the present discussion. 

\subsection{Ambiguities in non-Abelian gauge theory}
In non-Abelian gauge theories, there is more than one natural way to associate a spatial region \s\ with a part of the whole Hilbert space. Non-Abelian gauge theories have non-local degrees of freedom associated to Wilson loops.
 In the case of an Abelian gauge field, large Wilson loops are equivalent to integrals of local operators, so we do not need to consider them separately. For non-Abelian theories, however, a large Wilson loop cannot be rewritten as a product of local, gauge-invariant operators.\footnote{In conformal field theories, products of operators inserted at different points can be replaced by their OPE. Similarly one can attempt to construct OPE's for Wilson loops as in \cite{Mald}. However, an OPE does not converge if other operators are inserted within its radius of convergence, so in general Wilson loops cannot be replaced by local operators.}
 
Because the gauge-invariant degrees of freedom include nonlocal operators, we have to decide what to do with these nonlocal excitations when we define the Hilbert space associated with a subregion \reg. One natural choice, the bigger definition of \hrb, is to include in \hrb\ open Wilson loops that end on \s. With this definition, the full Hilbert space is a subset of $\hrb \otimes \mathcal{H}^B_{\reg^c}$, because both subsets allow open Wilson loops that end on \s, while in the full Hilbert space these loops must be tied together into a closed Wilson loop. An excellent discussion of these subtleties in the context of lattice gauge theory is given by Donnelly \cite{donnelly}. To describe this definition more precisely, \hrb\ is defined by all gauge field configurations in \reg\ modulo gauge transformations that are trivial on \s.

One could also make a different choice for how to treat Wilson loops  that do not fit inside \reg. The simplest possibility is not to allow Wilson loops to end on \s; in other words, \hrs\ is defined to be gauge field configurations in \reg\ modulo all gauge transformations, with no requirement that the gauge transformations should act trivially on \s. With this definition, the full Hilbert space of the theory is not contained in $\hrs \otimes \mathcal{H}^S_{\reg^c}$; the full Hilbert space also contains extra degrees of freedom associated to Wilson loops that do not fit in either side.

Since we are already thinking about what to do with large Wilson loops, there is another category of Wilson loops that could also be excluded from the Hilbert space associated with a region. These are Wilson loops that fit in \reg, but cannot in principle be measured within the causal diamond associated with \reg. In other words, these are loops with the property that no point within the causal diamond contains the loop in its backward lightcone. For spherical surfaces there are no such loops, but in general there are; for example, the strip contains long Wilson loops that are not in the backward lightcone of a single observer.

To summarize the above discussion, in non-Abelian gauge theories, there are several simple candidates for how to associate a spatial region to a subfactor of the Hilbert space. Each of these definitions has an associated density matrix and entanglement entropy that may violate strong subadditivity or other properties a Von Neumann entropy would have.

\subsection{Phenomenological suggestions for the CFT dual}

Having identified several ways to define the degrees of freedom associated to a region, the natural question is whether the causal holographic information is the entanglement entropy, with one of these choices? On the positive side, the arguments above that $\chi$ cannot be an entanglement entropy are evaded if there are degrees of freedom that are not contained in \hrs\ or $\mathcal{H}^S_{\reg^c}$. For example, it is no longer necessary that if the full quantum state is pure, then $\chi_\reg = \chi_{\reg^c}$, because the extra degrees of freedom can be entangled differently with \reg\ than its complement. A similar argument applies for the subadditivity condition.

However, there is a reason that the above definitions of the ``small" Hilbert space associated to a region are still not consistent with the geometric definition of $\chi$: the ``teleological" behavior of $\chi$ and the vanishing of the ``mutual causal information" \cite{Hubeny:2013hz}.  Suppose we start with the vacuum state and add a source to the future of the surface \reg\ but inside the associated causal diamond.  By construction, the source only changes the state inside its future lightcone, so it does not affect the state on \reg.  As long as the Hilbert space \hrs\ is defined in a state-independent way, as it is in all of the suggestions above, then the entanglement entropy of \reg\ must be independent of the source, because by construction it does not affect the density matrix on \reg. 
But one can see from the bulk definition that such a source does in general change the area $\chi.$

To identify a quantity that is also consistent with the teleological nature of the causal holographic information, we can consider something that is also well-motivated from AdS/CFT. The bulk causal wedge defined by \cite{hr} is the region that can be probed from the boundary diamond. Probing the bulk can be thought of as sending in a probe from some boundary point and then getting a signal back to the boundary. Therefore, the bulk causal wedge is the region of the bulk that should be encoded in the correlation functions of local, gauge invariant operators within the boundary diamond. 
Roughly, what we want to do is to do a coarse-graining where we identify states that cannot be distinguished by correlators of local, gauge-invariant operators. Such a definition does not seem to naturally lead to a tensor factor of the Hilbert space, and an associated density matrix, so we need a different construction. 

An option is that causal holographic information is the Von Neumann entropy for a density operator $\tilde{\rho}$ that is constructed from $\rho$ by a linear transformation. Linear maps $\rho \rightarrow \tilde{\rho}$ that are trace-preserving and completely positive can be written as \cite{book}:
\begin{equation} \label{kraus}
 \begin{aligned}
  \tilde{\rho} &= \sum_i M_i \rho M_i^{\dagger},
 \end{aligned}
\end{equation}
where the $M_i$ satisfy $\sum_i M_i^{\dagger}M_i = \id$.  One interesting case is a complete set of projection operators $\proj_i$ that add up to the identity:
\begin{equation} \label{project}
 \begin{aligned}
\sum_i \proj_i = \id,
 \end{aligned}
\end{equation}
and given the full density matrix $\rho$, a density matrix $\tilde{\rho}$ can be defined by
\be
\tilde{\rho} = \sum_i \proj_i \rho \proj_i.
\label{eq-cgrho}
\ee
This procedure ensures that the Von Neumann entropy of $\tilde{\rho}$ is greater than the Von Neumann entropy of $\rho$ \cite{book}, and this fits with the observation that the area of the causal information surface always has greater area than the extremal surface. 

A second interesting implementation of a linear map (\ref{kraus}), which we do not use here, is to consider a set of unitary operators:
\begin{equation}
 \begin{aligned}
  \tilde{\rho} &= \frac{1}{n}\sum_i^n U_i \rho U_i^{\dagger}.
 \end{aligned}
\end{equation}
This ensures that $S(\tilde{\rho}) \geq S(\rho)$ by concavity of Von Neumann entropy \cite{book}.

\paragraph{Proposal for simple causal diamonds.} Suppose the boundary region \reg\ corresponds to a simple causal diamond, defined by one future caustic and one past caustic. Then there is a natural choice for  the projection. Typically one can think of the projectors that project onto the eigenspaces of a particular hermitian operator. Since we want an operator associated to the causal development of the region \reg, and we do not want it to depend on details of the theory, a natural choice is the time evolution operator $U$ that evolves the state from the ``bottom" to the ``top" of the causal diamond. The evolution operator $U$ is unitary, so it can be written as the exponential of a Hermitian operator $A$, $U = \exp(i A)$. The eigenvectors of $A$ naturally pick out a set of projection operators that can be used in the above construction (\ref{eq-cgrho}).

There is one more encouraging observation that this construction may be on the right track, in addition to the above observation that the coarse-grained entanglement entropy is larger. \cite{hr} has analyzed in which cases $\chi = S$, concluding that the two quantities agree whenever the entanglement entropy can be thought of as a thermal entropy. In this special case,  the density matrix is already diagonal in the basis picked out by this Hamiltonian, so the coarse-graining (\ref{eq-cgrho}) has no effect.  (A subtlety is that the evolution from the bottom to the top requires $t \to \infty$.)

\paragraph{More complicated regions.}
The simplest boundary causal diamond is one with just one future caustic and one past caustic, so that in flat space, \s\ is a sphere. However, for most choices of \reg, the boundary region will be such that both $\mathcal{C}^+$ and $\mathcal{C}^-$ contain more than one point. In this case, there is another natural procedure for throwing away nonlocal information\footnote{This idea is due to reference \cite{bcc}; we thank Jan de Boer, Borun Chowdury, and Bartek Czech for discussions.}. Starting from the full density matrix $\rho_{\Sigma}$, define $\tilde \rho_{\Sigma}$ to be the maximum entropy density matrix that correctly reproduces the measurements contained in $any$ causal diamond contained within the boundary region. 

This procedure is motivated by considering which bulk region should be described. It has the advantage that it defines a $\tilde \rho$ that agrees with another observation about $\chi$. Suppose $A$ and $B$ are two spacelike patches on a Cauchy-surface on the boundary and $A \cap B = 0$, then $\chi_A+\chi_B-\chi_{A\cup B} =0$. This can be arranged iff the causal mutual information vanishes: $\tilde{\rho}_{A \cup B} = \tilde{\rho}_A \otimes \tilde{\rho}_B$. In this special case, one can show that the above procedure indeed leads to a density matrix that is a product.

To conclude this section of speculations, there are a couple of natural procedures for throwing away certain information from the density matrix. We hope that a combination of these provide a CFT definition of the causal holographic information. Several more detailed checks are possible that should allow one to either nail down the details of the construction, or show that it is the wrong idea.

\section{Conclusions}
We investigated properties of causal holographic information and in particular the coefficient of the logarithmically diverging piece. Using the work of Schwimmer and Theisen, we showed that this coefficient must be universal: state and regulator independent. Furthermore, the coefficient is nonlocal and constitutes a new conformal invariant. 

We captured the nonlocal behavior by introducing a function $\tau$ on the boundary surface $\partial \Sigma$, where $\tau$ can be thought of as the local ``height'' of the causal diamond. Using this function $\tau$, we derived an expression for the  universal log-term. An analysis of causal holographic information with a Gauss Bonnet dual enables a separation of the log-term coefficients in terms of the two central charges. 

Finally, we discussed properties of causal holographic information in relation to a possible the CFT-quantity that it represents.  We discussed several different operations on the full density matrix, with the hope that the causal holographic information is the Von Neumann entropy of the new density matrix. We gave some evidence indicating that our quantity agrees with the observed geometric properties of $\chi$.

\section*{Acknowledgements}
We thank Jan de Boer, Borun Chowdury, Bartek Czech, Veronika Hubeny, Matt Lippert, Matthew Headrick, Rob Myers, Mukund Rangamani, Matthew Kleban, and Jan Pieter van der Schaar for helpful discussions.

\appendix
\appendixpage
\addappheadtotoc

\section{Proof of formula for constant time slice}\label{tauproof}
In this appendix we will derive the near boundary expansion of the embedding function of the causal information surface (\ref{cis}) for an entanglement surface $\partial \Sigma$ that lies on a constant time slice on a flat boundary, using a pure AdS bulk metric:
\begin{equation}
 \begin{aligned}
ds^2 &= \frac{dz^2}{z^2}+\frac{1}{z^2}\eta_{\mu \nu}dx^{\mu}dx^{\nu}
 \end{aligned}
\end{equation}

For the boundary coordinates we will use the following notations interchangeably:
\begin{equation}
 \begin{aligned}
 x^{\mu} &= (t,\vec{x}). 
 \end{aligned}
\end{equation}

For any point $x^{\mu}(\xi_1,\xi_2)$ on $\partial \Sigma$ there is a unique point on top- and on the bottom of the causal diamond, $x_{\wedge}(\xi)$ on $\mathcal{C}^+$ and $x_{\vee}(\xi)$ on $\mathcal{C}^-$ respectively. These points are separated from the base point $x^{\mu}(\xi)$ by boundary null geodesics that are normal to $\partial \Sigma$ in $x^{\mu}(\xi)$. The causal diamond $\Diamond_{\Sigma}$ is symmetric in the $t=0$ plane, so if the future caustic $x_{\wedge}(\xi)$ is in the $t = \tau$ plane, the past caustic $x_{\vee}(\xi)$ is on the $t = -\tau$ plane. Since $\hat{t}$ is normal to $\partial \Sigma$, we can take the second normal vector $\hat{n}(\xi)$ to be in the $t=0$ plane. This allows us to express $ x^{\mu}_{\wedge}(\xi)$ in terms of $\hat{n}(\xi)$ and $\tau(\xi)$,
\begin{equation}\label{generate}
 \begin{aligned}
 x^{\mu}_{\wedge}(\xi) 	&= \left(x^0_{\wedge}(\xi),\vec{x}_{\wedge}(\xi)\right) \\
			&= \left(\tau(\xi), \vec{x}_{\partial \Sigma}(\xi)+\tau(\xi) \hat{n}(\xi)\right) \\
 \vec{x}_{\wedge}(\xi) 	&= \vec{x}_{\partial \Sigma}(\xi)+ \tau(\xi) \hat{n}(\xi),
 \end{aligned}
\end{equation}
where $\hat{n}(\xi)$ is the unit normal vector in the ($t=0$)-plane and $\tau(\xi)$ depends (in general nonlocally) on the geometry of $\partial \Sigma$. From now on we will not explicitly write that these are functions of $\xi$.

We parametrize the embedding function $x^M = (z,x^{\mu}_{bulk})$ of the causal information surface (\ref{cis}) with $z$ and $\xi^{\alpha}$ ($\alpha = 1,2$). Given a point $x^M = (z,x^{\mu}_{bulk})$ on the causal information surface we would like to identify the point(s) $\vec{x}(\xi)$ on $\partial \Sigma$ such that a past directed bulk lightray emanating from $x^{\mu}_{\wedge}(\xi)$ and a future directed bulk lightray $x^{\mu}_{\vee}(\xi)$ intersect at $x^M = (z,x^{\mu}_{bulk})$. For small $z$, $x^{\mu}_{bulk}$ and a particular $x^{\mu}(\xi)$ are close together ($O(z^2)$). Since the causal information surface consists of points that are as far into the bulk as possible while still being intersected by at least one null ray emanating from $\mathcal{C}^+$ and $\mathcal{C}^-$. The point $\vec{x}(\xi)$ that generates the caustics (\ref{generate}) from which the bulk point $x^M = (z,x^{\mu}_{bulk})$ can be reached by null rays should be such that the distance between $x^{\mu}_{bulk}$ and $x^{\mu}_{\wedge}(\xi)$ is minimal. The set of future caustics $\mathcal{C}^+$ is piecewise smooth and we construct the normal plane for a point $x^{\mu}_{\wedge}(\xi)$ in $\mathcal{C}^+$. The point $\vec{x}(\xi)$ should be such that:
\begin{itemize}\label{conditions}
\item{$x^{\mu}_{bulk}$ lies in the intersection of the normal plane of $x^{\mu}_{\wedge}(\xi)$ and the $t=0$ plane}
\item{$|x_{bulk}-x_{\wedge}(\xi)|^2 = -z^2$}.
\end{itemize}

One can construct two tangent vectors in $x_{\wedge}(\xi)$ (or $x_{\vee}(\xi)$) by taking derivatives with respect to the parameters $\xi^{\alpha}, \alpha = 1,2$,
\begin{equation}
 \begin{aligned}
 S^{\mu}_{\alpha \wedge} &= \partial_{\alpha}x^{\mu}_{\wedge} \\
 &= (S^{0}_{\alpha \wedge}, \vec{S}_{\alpha \wedge}) \\
 &= (\partial_{\alpha}\tau, \partial_{\alpha} \vec{x}_{\wedge}) \\
 \vec{S}_{\alpha \wedge} &= \partial_{\alpha} \vec{x}_{\wedge} \\
 &= \partial_{\alpha}\vec{x}_{\partial \Sigma}+(\partial_{\alpha}\tau)\hat{n} + \tau \partial_{\alpha}\hat{n} \\
 &= \vec{T}_{\alpha} + (\partial_{\alpha}\tau)\hat{n} + \tau \partial_{\alpha}\hat{n}
 \end{aligned}
\end{equation}
where $ \vec{T}_{\alpha} = \partial_{\alpha}\vec{x}_{\partial \Sigma}$.
One can construct $S_{\alpha \vee}$ similarly.

The space orthogonal to the tangent space $ span \left\lbrace  S^{\mu}_{\alpha \wedge} \right\rbrace$ in $x_{\wedge} \in \mathcal{C}^+$ is (in the most general case) two-dimensional and contains $(1,\hat{n})$. It is spanned by $(1,\hat{n})$ and by a second linearly independent vector. We construct such a vector $V^{\mu}_{\wedge}$ by demanding $V_{\wedge} \cdot S_{\alpha \wedge} = 0$ for $\alpha = 1,2$.

\begin{equation} \label{bwedge}
 \begin{aligned}
 V^{\mu}_{\wedge} &= (1,b^{\alpha}) \ \ \ \ (\text{Ansatz}) \\
 V_{\wedge}\cdot S_{\alpha \wedge} &= 0 \\
 &= -\partial_{\alpha}\tau + b^{\beta}\vec{T}_{\beta}\vec{T}_{\alpha}+\tau b^{\beta} \vec{T}_{\beta}\cdot \partial_{\alpha}\hat{n} \\
 &= -\partial_{\alpha}\tau + b^{\beta}\tilde{g}_{\beta \alpha}+ \tau b^{\beta}K_{\alpha \beta}
 \end{aligned}
\end{equation}
where $\tilde{g}_{\alpha \beta} = \vec{T}_{\alpha}\cdot \vec{T}_{\beta}$ is the induced metric on $\partial \Sigma$.

Equation (\ref{bwedge}) is solved by
\begin{equation}
 \begin{aligned}
 b_{\alpha} &= \frac{(\partial_{\alpha}\tau)(1+\tau K)-\tau K_{\alpha}^{\beta}(\partial_\beta\tau)}{1+\tau K +\frac{\tau^2}{2}(K^2-K_{\gamma \delta}K^{\gamma \delta})}.
 \end{aligned}
\end{equation}

Intersecting the $t=0$ plane with the surface spanned by $(1,\hat{n}(\xi))$ and $V(\xi)$ identifies the point $\vec{x}(\xi)$ together with the condition that $|x_{bulk} - x_{\wedge}(\xi)|^2 = -z^2$.

\begin{equation}
 \begin{aligned}
 x^{\mu}_{bulk} &= x^{\mu}_{\partial \Sigma} + a(0,\hat{n}-b^{\alpha}\vec{T}_{\alpha})^{\mu} \\
 x^{\mu}_{\wedge} &= x^{\mu}_{\partial \Sigma} + \tau (1,\hat{n})^{\mu} \\
 |x^{\mu}_{bulk} - x^{\mu}_{\wedge}|^2 &= -z^2
 \\
 &= -\tau^2+(a-\tau)^2+b^{\alpha}b^{\beta}\tilde{g}_{\alpha \beta} \\
 &= a^2(1+b^2)-2a\tau.
 \end{aligned}
\end{equation}
Solving for $a$ gives:

\begin{equation}
 \begin{aligned}
 a &= \frac{\tau}{1+b^2} \pm \frac{\tau}{1+b^2}\sqrt{1-\frac{z^2(1+b^2)}{\tau^2}} \\
 &= \frac{z^2}{2\tau} + O(z^4) \ \ (\text{Taking the minus solution}).
 \end{aligned}
\end{equation}
Giving the near boundary expansion of the embedding function of the causal information surface:
\begin{equation}
 \begin{aligned}
 \vec{x}_{bulk}(\xi,z) &= \vec{x}_{\partial \Sigma}(\xi) + \frac{z^2}{2\tau (\xi)}\hat{n}(\xi) - \frac{z^2}{2 \tau (\xi)}b^{\alpha}\vec{T}_{\alpha}(\xi)+O(z^4).
 \end{aligned}
\end{equation}

\section{Proof of the formula for a flat boundary}\label{lambdaproof}
Allowing the entanglement surface $\partial \Sigma$ to be a general spacelike surface in the Minkowski background and a pure AdS dual removes the mirror symmetry of the causal diamond in the $t=0$ plane we used in Appendix \ref{tauproof}. Given a point $(z,x^{\mu}_{bulk})$ on the causal information surface, the past and future caustics from which the bulk null geodesics intersect at $(z,x^{\mu}_{bulk})$ might be related to different points on $\partial \Sigma$.

Now we have to find two points $x^{\mu}_t = x^{\mu}(\xi^{\alpha})$ and $x^{\mu}_b = x^{\mu}(\xi^{\alpha}+\Delta^{\alpha})$ on $\partial \Sigma$ such that:
\begin{itemize}\label{conditions2}
\item{$x^{\mu}_{bulk}$ lies in the plane through $x^{\mu}_{\wedge}(\xi)$ orthogonal to $\mathcal{C}^+$}
\item{$x^{\mu}_{bulk}$ lies in the plane through $x^{\mu}_{\vee}(\xi+\Delta)$ orthogonal to $\mathcal{C}^-$}
\item{$|x_{bulk}-x_{\wedge}(\xi)|^2 = -z^2$}
\item{$|x_{bulk}-x_{\vee}(\xi+\Delta)|^2 = -z^2$},
\end{itemize} using the definition (\ref{lambs}).

From now we will interchangeably use the subscript $t$ for quantities that are evaluated at $\xi$ and the subscript $b$ for quantities that are evaluated at $\xi+\Delta$. So (\ref{lambs}) can also be written as:
\begin{equation}
 \begin{aligned}
x^{\mu}_{\wedge} &= x^{\mu}_t + \lambda_{\uparrow t}N^{\mu}_{\uparrow t} \\
x^{\mu}_{\vee} &= x^{\mu}_b + \lambda_{\downarrow b}N^{\mu}_{\downarrow b}.
 \end{aligned}
\end{equation}

In both $x_t$ and $x_b$ the two null normal vectors together with the two tangent vectors $T^{\mu}_{\alpha} = \partial_{\alpha}x^{\mu}$ constitute a basis and we can express $x^{\mu}_{bulk}$ in terms of these vectors:
\begin{equation}\label{expansion}
 \begin{aligned}
x^{\mu}_{bulk} &= x^{\mu}_t+\alpha_{\uparrow t}N^{\mu}_{\uparrow t}+ \alpha_{\downarrow t}N^{\mu}_{\downarrow t} + \alpha_{\downarrow t}b^{\alpha}_t T^{\mu}_{\alpha t} \\
&= x^{\mu}_b + \alpha_{\uparrow b}N^{\mu}_{\uparrow b}+\alpha_{\downarrow b}N^{\mu}_{\downarrow b} + \alpha_{\uparrow b}b^{\alpha}_b T^{\mu}_{\alpha b},
 \end{aligned}
\end{equation}
where $\alpha_{\uparrow t}$, $\alpha_{\uparrow b}$, $\alpha_{\downarrow t}$, $\alpha_{\downarrow b}$, $b^{\alpha}_t$ and $b^{\alpha}_b$ have to be determined by imposing the conditions listed above. 

Imposing the condition that $x^{\mu}_{bulk}$ lies in the plane through $x^{\mu}_{\wedge}(\xi)$ orthogonal to $\mathcal{C}^+$ is equivalent to extremizing $|x_{bulk}-x_{\wedge}|$ by varying $x_{t}$:
\begin{equation}
 \begin{aligned}
0 &= \partial_{\alpha}|x_{bulk}-x_{\wedge}|^2 \\
&= \partial_{\alpha}|x_{bulk}-x_{t}-\lambda_{\uparrow t}N_{\uparrow t}|^2 \\ 
&\Rightarrow (x^{\mu}_{bulk}-x^{\mu}_t)\eta_{\mu \nu}(T^{\nu}_{\alpha t}-\lambda_{\uparrow t}\partial_{\alpha}N^{\nu}_{\uparrow t}-(\partial_{\alpha}\lambda_{\uparrow t})N^{\mu}_{\alpha t}).
 \end{aligned}
\end{equation}

Similar for $|x_{bulk}-x_{\vee}|$.

Now using the expansion (\ref{expansion}) we find equations for $b^{\alpha}_t, \alpha =1,2$.
\begin{equation}\label{beq}
 \begin{aligned}
0 &= -(\partial_{\alpha}\lambda_{\uparrow t})+b^{\beta}_t\tilde{g}_{\alpha \beta}-b^{\beta}_t K_{\uparrow t \alpha \beta}\lambda_{\uparrow t},
 \end{aligned}
\end{equation}
where $\tilde{g}_{\alpha \beta} = \partial_{\alpha}x^{\mu}_t \partial_{\beta}x^{\nu}_t\eta_{\mu \nu}$.

This equation (\ref{beq}) (And similarly for $b_{\alpha b}$) is solved by:
\begin{equation}\label{bt}
 \begin{aligned}
b_{\alpha t} &= \frac{-\left(1+\lambda_{\uparrow t}K_{\uparrow t} \right)\partial_{\alpha}\lambda_{\uparrow t}-K_{\uparrow t \alpha}^{\beta}\partial_{\beta}\lambda_{\uparrow t}}{\left(1+\lambda_{\uparrow t}K_{\uparrow t}+\frac{\lambda_{\uparrow t}\lambda_{\uparrow t}}{2}\left(K_{\uparrow t}^2-K_{\uparrow t \gamma \delta}K_{\uparrow t}^{\gamma \delta} \right) \right)}.
 \end{aligned}
\end{equation}

Another condition can be obtained by demanding:
\begin{equation}\label{alpha1}
 \begin{aligned}
|x_{bulk}-x_{\wedge}|^2 &= -z^2 \\
&= 2(\alpha_{\uparrow t}-\lambda_{\uparrow t})\alpha_{\downarrow t}+ \alpha_{\downarrow t}^2b_{t}^2 \\
&\Rightarrow \alpha_{\downarrow t} = \frac{z^2}{2\lambda_{\uparrow t}}+O(z^4).
 \end{aligned}
\end{equation}
Similarly for $|x_{bulk}-x_{\vee}|$ giving a similar solution: 

\begin{equation}\label{alpha2}
 \begin{aligned}
\alpha_{\uparrow b} = \frac{z^2}{2\lambda_{\downarrow b}}+O(z^4).
 \end{aligned}
\end{equation}

Now using (\ref{alpha1}) and (\ref{alpha2}) and implicitly using (\ref{bt}) we can re-express (\ref{expansion}):
\begin{equation}
 \begin{aligned}
x^{\mu}_{bulk} &= x^{\mu}_t+\alpha_{\uparrow t}N^{\mu}_{\uparrow t}+ \frac{z^2}{2\lambda_{\uparrow t}}N^{\mu}_{\downarrow t} + \frac{z^2}{2\lambda_{\uparrow t}}b^{\alpha}_t T^{\mu}_{\alpha t}+... \\
&= x^{\mu}_b + \frac{z^2}{2\lambda_{\downarrow b}}N^{\mu}_{\uparrow t}+\alpha_{\downarrow b}N^{\mu}_{\downarrow b} + \frac{z^2}{2\lambda_{\downarrow b}}b^{\alpha}_b T^{\mu}_{\alpha b}.
 \end{aligned}
\end{equation}

Now we can expand in $\Delta$ using that $\Delta^{\alpha}\partial_{\alpha}x^{\mu}$ is of order $O(z^2)$:
\begin{equation}
 \begin{aligned}
x^{\mu}_{bulk} &= x^{\mu}+\alpha_{\uparrow t}N^{\mu}_{\uparrow}+ \frac{z^2}{2\lambda_{\uparrow}}N^{\mu}_{\downarrow } + \frac{z^2}{2\lambda_{\uparrow }}b^{\alpha}_t T^{\mu}_{\alpha}+... \\
&= x^{\mu} +\Delta^{\alpha}\partial_{\alpha}x^{\mu}+ \frac{z^2}{2\lambda_{\downarrow}}N^{\mu}_{\uparrow }+\alpha_{\downarrow b}N^{\mu}_{\downarrow} + \frac{z^2}{2\lambda_{\downarrow}}b^{\alpha}_b T^{\mu}_{\alpha}+...
 \end{aligned}
\end{equation}

We use linear independence to find equation for the coefficients of $T^{\mu}_{\alpha}$, $N^{\mu}_{\uparrow}$ and $N^{\mu}_{\downarrow}$:
\begin{equation}
 \begin{aligned}
\frac{z^2}{2\lambda_{\uparrow}}b^{\alpha}_t &= \Delta^{\alpha}+\frac{z^2}{2\lambda_{\downarrow}}b^{\alpha}_b+O(z^4) \\
\alpha_{\uparrow t} &= \frac{z^2}{2\lambda_{\downarrow}}+O(z^4) \\
\frac{z^2}{2\lambda_{\uparrow}} &= \alpha_{\downarrow b}+O(z^4).
 \end{aligned}
\end{equation}

Now we can expand the embedding function using $x_t = x(\xi)$:
\begin{equation}
 \begin{aligned}
 x^{\mu}_{bulk}(z,\xi) &= x^{\mu}(\xi) + \frac{z^2}{2}\left(\frac{N^{\mu}_{\uparrow}}{\lambda_{\downarrow}}+\frac{N^{\mu}_{\downarrow}}{\lambda_{\uparrow}} \right)+\frac{z^2}{2\lambda_{\uparrow}}b^{\alpha}_{t}T^{\mu}_{\alpha} + O(z^4). 
 \end{aligned}
\end{equation}

And for the relation between we find $x_t$ and $x_b$:

\begin{equation}
 \begin{aligned}
x^{\mu}_t &= x^{\mu}(\xi) \\
x^{\mu}_b &= x^{\mu}(\xi + \Delta) \\
&\Rightarrow \Delta^{\alpha} = \frac{z^2}{2}\left(\frac{b^{\alpha}_t}{\lambda_{\uparrow}}-\frac{b^{\alpha}_b}{\lambda_{\downarrow}} \right)+O(z^4).
 \end{aligned}
\end{equation}

\end{document}